\documentclass[aps,prb,amsmath,amssymb,footinbib,showpacs,twocolumn]{revtex4-2}
\usepackage{amsmath}
\usepackage{amssymb}
\usepackage{amsthm}
\usepackage{setspace}
\usepackage{graphicx}
\usepackage{braket}
\usepackage{mathrsfs}
\usepackage{physics}
\usepackage{float}
\usepackage[colorlinks = true,linkcolor = red,urlcolor  = blue,citecolor = blue,anchorcolor = blue]{hyperref}
\usepackage[utf8]{inputenc}
\usepackage[english]{babel}
\usepackage{bm}
\usepackage{xcolor}
\usepackage{ulem}
\usepackage[section]{placeins}

\begin{document}
	\title{Disorder effects on Majorana zero modes:  Kitaev chain versus semiconductor nanowire}
	\author{Haining Pan}
	\author{S. Das Sarma}
	\affiliation{Condensed Matter Theory Center and Joint Quantum Institute, Department of Physics, University of Maryland, College Park, Maryland 20742, USA}
	
	\begin{abstract}
		Majorana zero modes in a superconductor-semiconductor nanowire have been extensively studied during the past decade. Disorder remains a serious problem, preventing the definitive observation of topological Majorana bound states. Thus, it is worthwhile to revisit the simple model, the Kitaev chain, and study the effects of weak and strong disorder on the Kitaev chain. By comparing the role of disorder in a Kitaev chain with that in a nanowire, we find that disorder affects both systems but in a nonuniversal manner. In general, disorder has a much stronger effect on the nanowire than the Kitaev chain, particularly for weak to intermediate disorder. For strong disorder, both the Kitaev chain and nanowire manifest random featureless behavior due to universal Anderson localization. Only the vanishing and strong disorder regimes are thus universal, manifesting respectively topological superconductivity and Anderson localization, but the experimentally relevant intermediate disorder regime is nonuniversal with the details dependent on the disorder realization in the system.
	\end{abstract}
	
	\date{\rm\today}
	\maketitle
	
	\section{Introduction}\label{sec:introduction}
	 Majorana zero modes, which are neither fermions nor bosons and have many unconventional properties such as non-Abelian statistics, are regarded as a promising platform for error-free topological quantum computing~\cite{kitaev2003faulttolerant,dassarma2005topologically,nayak2008nonabelian}. In 2001, it was realized that Majorana zero modes can emerge as localized zero-energy bound states in an idealized model of a one-dimensional spinless $ p $-wave superconductor, which has since become known as a Kitaev chain~\cite{kitaev2001unpaired}. The Hamiltonian of a Kitaev chain is composed of a $ p$-wave superconducting pairing term, a nearest-neighbor hopping term, and an on-site chemical potential term. Later, several realistic proposals were made for the laboratory realization of effective spinless $ p $-wave superconductivity and the associated Majorana zero modes in different two- and one-dimensional (2D and 1D) systems~\cite{fu2008superconducting,zhang2008ip,lutchyn2010majorana,alicea2010majorana,sau2010generic,sau2010nonabelian,oreg2010helical}. Effectively spinless $ p $-wave superconductors carrying zero-energy Majorana bound states, the so-called Majorana zero modes, are topological superconductors. One of the most-studied proposals is the superconductor-semiconductor hybrid nanowire system~\cite{sau2010nonabelian,oreg2010helical}. The semiconductor requires a large Rashba-type spin-orbit coupling and Zeeman field to explicitly break the spin-rotational symmetry and time-reversal symmetry. The semiconductor also acquires an effective $ p $-wave superconducting pairing by a proximitized $ s $-wave superconductor in the presence of Rashba spin-orbit coupling and Zeeman spin splitting.  The possibility of topological superconductivity in such semiconductor nanowires, with Majorana zero modes localized at the wire ends, has been extensively studied theoretically and experimentally over the last 10 years with $ >1000 $ publications in the literature, although the actual existence of topological Majorana zero modes in nanowires has still not been established beyond a reasonable doubt.
	
	Two years after the original theoretical proposal, the first Majorana experiment on a superconductor-semiconductor nanowire~\cite{mourik2012signatures} appeared with the observation of zero-bias tunnel conductance peaks, considered to be a hallmark of Majorana zero modes. Although the maximal conductance of such zero-bias peaks was far below (by a factor of 10 or more) the quantized value of $ 2e^2/h $~\cite{sengupta2001midgap} in Ref.~\onlinecite{mourik2012signatures}, it inspired many follow-up experiments~\cite{das2012zerobias,deng2012anomalous,churchill2013superconductornanowire,finck2013anomalous, deng2016majorana,albrecht2016exponential,nichele2017scaling,zhang2017ballistic,chen2017experimental,vaitiekenas2018effective,moor2018electric,zhang2018quantizeda,zhang2021large,bommer2019spinorbit,grivnin2019concomitant,chen2019ubiquitous,anselmetti2019endtoend,menard2020conductancematrix,yu2021nonmajorana,puglia2020closing} producing better zero-bias conductance peaks by improving the quality of the samples.  Recently, a maximal conductance of almost $ 2e^2/h $, which is so far the closest one to the predicted Majorana quantized value, was reported in Refs.~\onlinecite{nichele2017scaling,zhang2018quantizeda,zhang2021large}. These reported  ``almost-quantized'' zero-bias conductance peaks are, however, neither very robust nor very stable as a function of system parameters (such as the temperature,  applied magnetic field, and gate voltage), casting doubts on their topological Majorana origin~\cite{zhang2021large}.
	
	Although the quality of the samples has greatly improved since the first experiment, there are several recent experiments~\cite{puglia2020closing,yu2021nonmajorana,chen2019ubiquitous} indicating that disorder is a serious problem. Because the disorder-induced Andreev bound states, which can mimic most signatures of the topological Majorana zero modes~\cite{pan2020physical}, are ubiquitous in the nanowire, it is worthwhile to revisit the original Kitaev chain and compare it with the nanowire, taking disorder into account equivalently in both systems, to better understand the role of disorder. The context of this in-depth study in the current paper is our recent finding~\cite{pan2020physical} that disorder, by itself, can generically give rise to ``trivial'' zero-bias tunnel conductance peaks which have conductance values around  $ 2e^2/h $ without the manifestation of any topological superconductivity.
	
	Previously, a disorder effect on the Kitaev chain and other associated quantum wires has been discussed in Refs.~\onlinecite{motrunich2001griffiths,brouwer2000localization,gruzberg2005localization,potter2010multichannel,brouwer2011probability,lobos2012interplay,degottardi2013majoranaa,degottardi2013majorana,hegde2016majorana}, and a disorder effect on the superconductor-semiconductor nanowire was also intensively studied in Refs.~\onlinecite{brouwer2011topological,lutchyn2011search,akhmerov2011quantized,sau2012experimental,liu2012zerobias,hui2015bulk,sau2013density,liu2017andreev,haim2019benefits,pan2020physical,pan2021threeterminal}. But there has been no study comparing disorder effects on a Kitaev chain and semiconductor nanowire using equivalent models treating both on an equal footing, perhaps because the Kitaev chain is studied entirely from a formal theoretical perspective (as it has no experimental relevance) whereas most nanowire disorder studies focus on understanding the experimentally observed zero-bias conductance peaks.  In the current paper, we bridge this gap, focusing on disorder effects equivalently in both systems for a deeper understanding of the universal and nonuniversal aspects of disorder on topological superconductivity in one-dimensional systems.  While the universal aspects of disorder may be theoretically more interesting, the existing experimental studies are likely to be dominated by nonuniversal aspects as indicated by substantial sample-to-sample variations in the observed tunneling spectroscopy data on Majorana experiments. Thus, we want to connect disorder effects on these two systems, and compare their differences.
	
	We first calibrate the Kitaev chain and superconductor-semiconductor nanowire to ensure that the effective lengths of the two systems are qualitatively the same so that they are topologically comparable. Then we introduce an uncorrelated Gaussian distribution of on-site disorder to both systems and exactly calculate the local density of states (LDOS). By making a qualitative one-to-one mapping, we can then directly compare disorder effects on the two systems. We find that, in general, the nanowire is more susceptible to disorder than the Kitaev chain. We analyze the likelihood of the occurrence of disorder-induced trivial zero-energy bound states statistically, and find that trivial zero-energy bound states are typically much more prominent in the nanowire than in the Kitaev chain. This relative immunity of the Kitaev chain to disorder compared with nanowires may be a reason why the importance of disorder in Majorana experiments was under appreciated until recently, since most theoretical analyses tend to focus on the universal aspects of disorder for which the simplified Kitaev model is adequate.  Our work shows that disorder effects are in fact significantly different in the two systems in spite of both systems belonging to class D or BDI in the topological classification.
	
	The remainder of this paper is organized as follows. In Sec.~\ref{sec:theory}, we review the Hamiltonian of a Kitaev chain and nanowire in its standard form, respectively, and explain how disorder is introduced in the model, and how the topological quantum phase transition (TQPT) is defined in a finite system in the presence of disorder. We also clarify and justify our choice of parameters for these two distinct systems to ensure that they are comparable. In Sec.~\ref{sec:results}, we present our main results, which include a short Kitaev chain in class BDI and class D, a long Kitaev chain in class BDI, as well as a short and long nanowire. All of these results are thoroughly studied in the presence of disorder varying from weak to strong. In Sec.~\ref{sec:discussion}, we statistically estimate the likelihood of the occurrence of disorder-induced trivial zero-energy bound states, and compare the role of disorder in the two systems. We present our conclusion in Sec.~\ref{sec:conclusion}. In the Appendix, we show the calculated topological invariant for the nanowire and the Kitaev chain, respectively, to complement the disorder results presented in the main text.
	\section{Theory}\label{sec:theory}
	
	\subsection{Hamiltonian of a Kitaev chain}
	The Hamiltonian of a Kitaev chain of length $ L $ is composed of a $ p $-wave superconducting pairing term $ \Delta $, a nearest-neighbor hopping term $ t $, and an on-site chemical potential term $ \mu_i $~\cite{kitaev2001unpaired},
	\begin{eqnarray}\label{eq:kitaev}
		H_{\text{Kitaev}}&=&\sum_{i=1}^{L-1} \qty(-t c_{i+1}^\dagger c_i+\Delta c_{i+1}^\dagger c_i^\dagger+\text{h.c.})\nonumber\\
		&-&\sum_{i=1}^{L} \mu_i c_i^\dagger c_i,
	\end{eqnarray}
	where $ c_i^\dagger $ ($ c_{i} $) creates (annihilates) an electron at site $ i $, and $\Delta=\abs{\Delta} e^{i\theta} $. 
	
	We will study two ensembles: (a) class BDI with the choice of $ \theta=0 $; (b) class D with the choice of $ \theta=\frac{\pi}{2} $. Here, the Kitaev chain in class BDI preserves the time-reversal symmetry, which requires $ \theta $ to be 0 or $ \pi $. Therefore, we simply set it to zero following the original choice in Ref.~\onlinecite{kitaev2001unpaired}. 
	
	However, the Kitaev chain in class D breaks the time-reversal symmetry, which requires any value of $ \theta $ other than 0 or $ \pi $. Therefore, the choice of $ \theta=\frac{\pi}{2} $ in class D is rather arbitrary, and we are just using this specific value of $ \theta $ to do the calculation without loss of generality. (We have also verified that any other value of $ \theta $  leads to the same results as that in $ \theta=\frac{\pi}{2}  $.) In fact, the phase of pairing $ \Delta $ will not affect the results since it can be absorbed into the Majorana operators~\cite{kitaev2001unpaired}. But we emphasize this consistency explicitly by directly presenting and comparing their results.
	
	In the Bogoliubov–de Gennes (BdG) formalism, the Hamiltonian~\eqref{eq:kitaev} is transformed into $ H=\frac12 C^\dagger H_{\text{BdG}} C $, where  $ C=\qty(c_1,\dots,c_L,c_1^\dagger,\dots,c_L^\dagger)^T $ , 
	\begin{eqnarray}\label{eq:bdg}
		H_{\text{BdG}}&=&-\sum_{i=1}^{L-1}\qty[\qty(t\tau_z+i\Delta\tau_y)\ketbra{i}{i+1}+\text{h.c.}]\nonumber\\
		&-&\sum_{i=1}^{L} \mu_i \tau_z \ketbra{i}{i},
	\end{eqnarray}

	and $ \vec{\tau} $ is the Pauli matrix that acts on the particle-hole space.
	\subsection{Hamiltonian of nanowire}
	
	The Hamiltonian of the superconductor-semiconductor nanowire of length $ L $ includes an effective $ p $-wave superconducting pairing term $ \Delta $ proximitized by an $ s $-wave superconductor, a chemical potential term $ \mu $ in the semiconductor, a Zeeman energy term $ V_Z $ arising from the magnetic field, and a Rashba-type spin-orbit coupling term $ \alpha $~\cite{sau2010nonabelian},
	\begin{eqnarray}\label{eq:nanowire}
		H_{\text{nanowire}}&=&\frac12\int_{0}^{L} dx ~\hat{\Psi}^\dagger(x) \left[\left(\-\frac{\hbar^2\partial^2_x}{2m^*}  -i \alpha \partial_x \sigma_y - \mu \right)\tau_z \right. \nonumber\\
		&+& \left. V_Z\sigma_x + \Delta \tau_x \vphantom{\frac12} \right] \hat{\Psi}(x),
	\end{eqnarray}
	where $\hat{\Psi}(x)=\left[\hat{\psi}_{\uparrow}(x),\hat{\psi}_{\downarrow}(x),\hat{\psi}_{\downarrow}^\dagger(x),-\hat{\psi}_{\uparrow}^\dagger(x)\right]^T$ represents a position-dependent spinor, $\vec{\bm{\sigma}}$ and $\vec{\bm{\tau}}$ are the Pauli matrices that act on the spin space and particle-hole space respectively, and $ m^* $ is the effective mass. The magnetic field, which contributes to the Zeeman energy $ V_Z $, is applied along the longitudinal direction of the nanowire, and also perpendicular to the direction of the Rashba-type spin-orbit coupling. 
		
	Using the convention of a Nambu spinor, we can transform Eq.~\eqref{eq:nanowire} into a BdG Hamiltonian. Then, replacing the differential operator with the finite difference,  we can further transform the continuum Hamiltonian into a tight-binding Hamiltonian~\cite{dassarma2016how},
	\begin{eqnarray}\label{eq:nw}
	&	H_{\text{TB}} &=\sum_{i=1}^{N-1} \qty[-t \ketbra{i+1}{i}\tau_z + i\alpha_R \ketbra{i+1}{i} \sigma_y\tau_z+\text{h.c.}] \nonumber \\		
		&+&\sum_{i=1}^{N} \qty[\Delta\ketbra{i}{i} \tau_x + \qty(2t-\mu_i)\ketbra{i}{i} \tau_z + V_Z\ketbra{i}{i}\sigma_x],
	\end{eqnarray}
	where we use the fictitious lattice constant  $ a $  to discretize the continuum Hamiltonian, and thus $ L=aN $,  $ t=\frac{\hbar^2}{2m a^2} $, and $ \alpha_R=\frac{\alpha}{2a} $. Here, we write down the chemical potential $ \mu $ in terms of position $ i $ explicitly because we will introduce the on-site disorder later, which would make the chemical potential essentially a random term.
	
	The nanowire BdG Hamiltonian~\eqref{eq:nw} now shares some similarities with the Kitaev chain BdG Hamiltonian~\eqref{eq:bdg}. We note that there is a major difference between the nanowire and Kitaev chain in the dimension of Hilbert space: The nanowire doubles the Hilbert space due to the spin degrees of freedom  whereas the Kitaev model is explicitly spinless by construction.
	
	 However, if we focus on their similarities, we can make a loose one-to-one mapping: (1) The hopping term $ t $ in the Kitaev chain maps to the Zeeman field $ V_Z $ in the nanowire; (2) the chemical potential and superconducting pairing term just map to themselves but in the other system. Although this one-to-one mapping is not rigorous formally, it makes a direct intuitive physical comparison possible, and thus is useful to help us understand the role of disorder in the two systems.
	
	\subsection{On-site disorder}
	
	To compare the role of disorder in the two systems, we introduce the on-site disorder to both $ \mu_i $ in the Kitaev chain in Eq.~\eqref{eq:bdg} and nanowire in Eq.~\eqref{eq:nw}. The random $ \mu_i $ is drawn from an uncorrelated Gaussian distribution with the mean value of $ \bar{\mu} $ and variance of $ \sigma_{\mu}$~\cite{pan2020physical,pan2021threeterminal}. Here, $ \bar{\mu} $ corresponds to the constant chemical potential in the pristine case (i.e., $ \sigma_{\mu}/\bar{\mu}=0 $). We present the results in the presence of disorder from a weak ($ \sigma_{\mu}/\bar{\mu}\sim0.4 $) up to a strong ($ \sigma_{\mu}/\bar{\mu}\sim 5$) level. For even weaker disorder, the results are essentially the same as in the well-known pristine situations. {We note that random disorder is present the nanowire, because the opposite scenario--- localized disorder in a subset of the wire--- can effectively break the wire into two (or multiple) wires, where each part can be described by Eq.~\eqref{eq:nanowire}, if the disorder is very strong. If the localized disorder is weak, the physics and the topology of the wire will not be much affected.}

	\subsection{Topological invariant}
	
	Because we are studying finite systems, it is unclear how to define the TQPT, which is a property of the thermodynamic limit. In principle, there is no TQPT in the finite system since the original topological invariant is only well defined in the infinite system and all finite systems should be essentially trivial. For example, the original TQPT, defined by the Pfaffian of the Kitaev chain Hamiltonian, measures the fermion parity switch~\cite{hegde2016majorana}. To show this, we numerically calculate the Pfaffians~\cite{wimmer2012algorithm} and present them in the Appendix. 
	
	To find an equivalent quantity in finite systems, we use the Lyapunov exponent (LE), which measures the inverse of the decay length of the ground state wave function in the finite Kitaev chain. The LE is defined based on the larger eigenvalue of the transfer matrix~\cite{hegde2016majorana,motrunich2001griffiths,degottardi2011topological,degottardi2013majorana,degottardi2013majoranaa}, $ \gamma=\lim\limits_{L\rightarrow\infty}\log\qty[\max(\abs{\lambda})] $, where $ \lambda $ is the set of eigenvalues of the Majorana transfer matrix $ A $ of the Kitaev chain. The Majorana transfer matrix is defined as $ A=\prod_i A_i $, where 
	\begin{equation}\label{key}
		A_i=\mqty(-\frac{\mu_i}{t+\Delta} & -\frac{t-\Delta}{t+\Delta}\\ 1 & 0).
	\end{equation}
	If the LE is positive, it indicates an increasing wave function from the end to the bulk of the nanowire, and thus it represents an extended ground state which corresponds to the trivial phase with no Majorana zero modes. However, when the LE becomes negative, it indicates an exponentially decaying wave function from the end of the nanowire to the bulk, and thus it represents a localized state which corresponds to the topological Majorana zero modes. Therefore, the vanishing (change of sign) of LE will tell us where the TQPT occurs in the finite Kitaev chain. This definition is unique and useful for the finite chain where the use of the Pfaffian becomes problematic.  For a long enough system size, the two indicators give the same result. 
	
	We can use a similar topological invariant--- topological visibility (TV)~\cite{fulga2011scattering,akhmerov2011quantized,dassarma2016how,pan2020physical}--- to define the TQPT in the finite nanowire. The TV is defined as $ Q=\det(r) $, which is also constructed based on the LE~\cite{akhmerov2011quantized}. Here, $ r $ is the reflection block of the $ S $ -matrix at one end of the nanowire (it does not matter which end since both ends will lead to the same result). The TV is positive in the trivial regime and negative is the topological regime. Thus, when it crosses zero, the TQPT occurs in the nanowire.
	
	We present the calculated Pfaffian and LE for Kitaev chains, and TV for nanowires in the Appendix. 	
	
	\subsection{Choice of parameters}
	
 	To make the two cases comparable, we need to choose the effective wire lengths for the two systems to be roughly the same. Therefore, for the nanowire, we choose a set of parameters of an InSb-Al hybrid nanowire~\cite{zhang2018quantizeda,zhang2021large} with a proximitized superconducting gap $ \Delta=0.2 $ meV, chemical potential $ \mu=1 $ meV, effective mass $ m^*= 0.015 m_e$ ($ m_e $ is the electron rest mass), and Rashba-type spin-orbit coupling $ \alpha=0.5$ eV \AA . These are the accepted parameter choices for simulating the existing InSb-Al experimental nanowires. 
 	
 	Similarly, we calibrate the Kitaev chain by setting the dimensionless chemical potential $ \mu=5 $ and $ p $-wave superconducting pairing gap $ \Delta=1 $, which are proportional to those in the nanowire. (The Kitaev chain being an idealized model, the parameter choice is dimensionless.)
 	
 	For the short wire, we choose the length to be $ L=3~\mu$ m. (Although $ L=3~\mu$ m is called ``short wire", it is already longer than the recent experimental devices~\cite{grivnin2019concomitant,vaitiekenas2018effective,zhang2018quantizeda,zhang2021large,moor2018electric,bommer2019spinorbit,nichele2017scaling}--- it is well known that the current experimental wire lengths are simply too short.) We then set the dimensionless number of sites $ L=20 $ for the Kitaev chain so that their effective lengths will qualitatively be the same ($ \sim 10 $ times of the coherence length at the TQPT in the pristine case). We note that what matters is the dimensionless wire length measured in units of the coherence length, which should, in principle, be much larger than unity (i.e., wire length $ \gg $ coherence length) for topological superconductivity with Majorana zero modes to occur. 
 	
	For the long wire, we choose the length to be $ L=10~\mu $ m for the nanowire, and the dimensionless number of sites $ L=50 $ for the Kitaev chain to ensure their effective dimensionless lengths in the two systems are of the same magnitude.

	\subsection{Local density of states}	
		
	Lastly, we discuss which theoretical quantity we should use to compare the two systems. In the nanowire experiment,~\cite{mourik2012signatures,das2012zerobias,deng2012anomalous,churchill2013superconductornanowire,finck2013anomalous, deng2016majorana,albrecht2016exponential,nichele2017scaling,zhang2017ballistic,chen2017experimental,vaitiekenas2018effective,moor2018electric,zhang2018quantizeda,zhang2021large,bommer2019spinorbit,grivnin2019concomitant,chen2019ubiquitous,anselmetti2019endtoend,menard2020conductancematrix,yu2021nonmajorana,puglia2020closing} what is being measured is the tunneling conductance at one end of the nanowire through an NS junction. Under the assumption of a single-band model, the local tunnel conductance varies between $ 0 $ and $ 4e^2/h $, where $ 2e^2/h $ is the quantized conductance of Majorana bound states (though the conductance of trivial Andreev bound states can also accidentally manifest a conductance of $ 2e^2/h $ due to disorder~\cite{liu2017andreev,pan2020generic,pan2020physical}). One theoretical quantity which is closely related to the tunnel conductance, in addition to being a meaningful characterization of both systems, is the local density of states (LDOS) at the wire ends, which contains information about the system wave function.
	
	Because the local conductance at one end of the nanowire probes the LDOS at that end, the LDOS at a particular end can qualitatively reflect the local conductance if measured from the same end. This is important in the attempt to make a direct comparison between the nanowire and Kitaev chain. The LDOS at energy $ \omega $ and position $ i $ of the system that is described by a tight-binding Hamiltonian $ H $ is defined as $ \text{LDOS}(\omega,i)=-\frac{1}{\pi} \qty{\Im\qty[\tr_{\sigma,\tau}({\omega+\delta_0-H})^{-1})]}_{i,i} $~\cite{huang2018quasiparticle}, where $ \tr_{\sigma,\tau} $ is a partial trace over the spin space $ \sigma $ and particle-hole space $ \tau $, $ \Im[\dots] $ takes the imaginary part, and $ \delta_0 $ is small for the inverse of lifetime.

	In the Kitaev chain, there is no direct experimental measurement of the tunneling transport because it is an idealized spinless model. Therefore, there is no experimental report of the ``local conductance" in the Kitaev chain similar to the nanowire. Of course, we can attach a ``lead" by introducing the hopping from the first site in the Kitaev chain to the neighboring site in the lead, and creating a potential barrier at the interface manually. But it is too artificial and unnecessary. Instead, similar to the nanowire, we can simply associate the LDOS at one end of the Kitaev chain with the ``local conductance" (if we could perform such a transport experiment) at the same end. The LDOS is theoretically a more useful quantity to look at since it is a property only of the system (the chain or the nanowire) and the properties of the lead do not enter into consideration.  In addition, the temperature and/or tunneling amplitude, which are extremely important in determining the tunnel conductance~\cite{sengupta2001midgap,setiawan2017electron}, do not contaminate the LDOS, which is a unique function only of the eigenstates of the system, and as such codify the topological properties rather uniquely.  Of course, LDOS itself should not be compared directly with conductance measurements except qualitatively, but our goal in the current work is a theoretical comparison between a Kitaev chain and nanowire, which is better done with LDOS and not conductance.
	
	Therefore, we will show all the results in terms of LDOS at the left end and the right end of the chain, which is an analog to the current transport experiment in the nanowire. We can compare the correlation of the LDOS from both ends in a similar way of analyzing the correlation between the left and right local conductance ($ G_{\text{LL}} $ and $ G_{\text{RR}} $)~\cite{menard2020conductancematrix,lai2019presence}. In this case, a peak of LDOS at zero energy will indicate a zero-energy (or near-zero-energy, within the energy resolution) bound state in the system (which can be trivial or topological depending on the parameter values). 
	
	Finally, we also present the density of states (DOS) in the system, which is a spatial average of LDOS along the wire, to show whether there is a bulk zero-energy state in the system (which can be an extended state or a localized state, i.e., whether or not the bulk gap closes at some parameter value as should happen at the TQPT~\cite{stanescu2012close}). We note that LDOS (DOS) signify states at the boundary (bulk) of the system, providing somewhat complementary information.

	\section{Results}\label{sec:results}
	In this section, we present the representative results in the short Kitaev chain in class BDI (Sec.~\ref{sec:BDI}) and class D (Sec.~\ref{sec:D}), the long Kitaev chain in class BDI (Sec.~\ref{sec:BDI50}), the short nanowire (Sec.~\ref{sec:nw}), and the long nanowire (Sec.~\ref{sec:nw1000}). In each disorder, we first present DOS, LDOS at the left end, and LDOS at the right end in the pristine case from the top to the bottom row. Although zero-disorder pristine results are all very well known, we are showing them for completeness so that it is convenient for readers to compare directly with the disordered cases. Then we will gradually increase disorder from weak ($ \sigma_\mu/\bar{\mu}=0.4 $) to intermediate ($ \sigma_\mu/\bar{\mu}=1 $) and finally to strong disorder ($ \sigma_\mu/\bar{\mu}=5 $). We note that the following results are all shown for one specific configuration of randomness without averaging over disorder. Results for other disorder configurations look statistically the same, but the details differ similar to the experimental sample-to-sample conductance data variations.

	\subsection{Short Kitaev chains in class BDI}\label{sec:BDI}
	
	 	\begin{figure*}[ht]
	\centering
	\includegraphics[width=\textwidth]{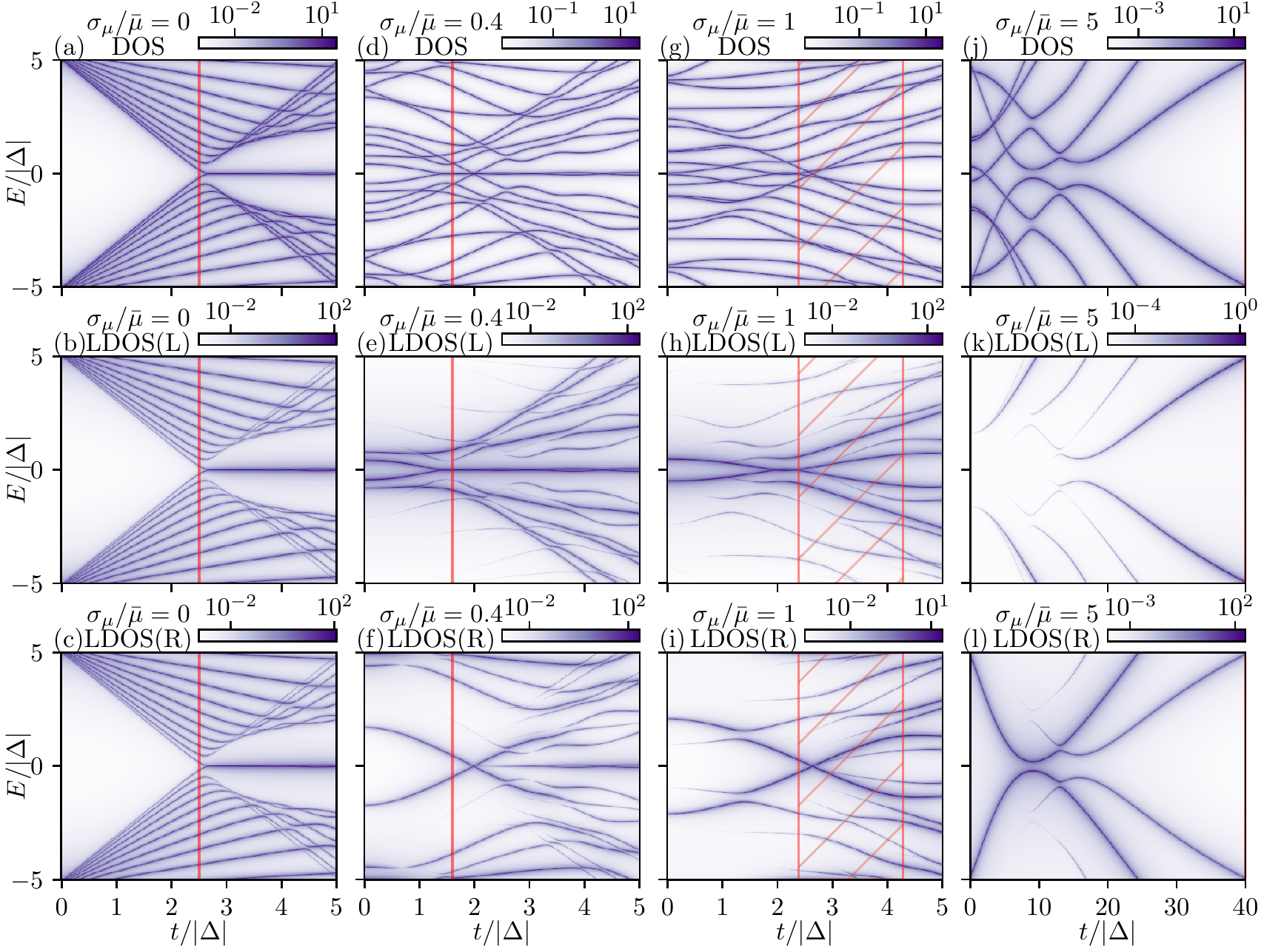}
	\caption{The representative results of a short Kitaev chain with $ L=20$ in class BDI ($ \theta=0 $). 
	(a)-(c) show the pristine wire; 
	(d)-(f) show a weak disorder with $ \sigma_\mu/\bar{\mu}=0.4 $; 
	(g)-(i) show an intermediate disorder with $ \sigma_\mu/\bar{\mu}=1 $;  
	(j)-(l) show a strong disorder with $ \sigma_\mu/\bar{\mu}=5 $. The first row represents the DOS; the second row represents the LDOS at the left end; the third row represents the LDOS at the right end. The red line indicates the TQPT, which is defined by the LE. The hatched region represents a transition from the trivial to the topological regime, between which the LE is unstable around zero. The corresponding LE and Pfaffian are shown in Fig.~\ref{fig:BDIPf}. Other parameters in the Kitaev chain are $ \bar{\mu}/\abs{\Delta}=5 $, and $ \abs{\Delta}=0.2 $.}
	\label{fig:BDI}
\end{figure*}

	We first present the results of the short Kitaev chains in class BDI, as shown in Fig.~\ref{fig:BDI}. The first column shows the results of the well-known pristine Kitaev chain. The TQPT occurs when $ t=\frac{\mu}{2} $, below which is the trivial regime. In the first row, the absence of zero-energy bound states in the trivial regime indicates that the system is manifesting the ``good'' (i.e., topological) Majorana zero modes~\cite{pan2020physical}. In the second and third rows, we present the LDOS at the left and right end, respectively, where we find they are correlated as expected as the Majorana modes must come in pairs.

	We also plot the calculated Pfaffian in Fig.~\ref{fig:BDIPf} in the Appendix (see the red line), where we find that the Pfaffian is +1 in the trivial regime. However, it quickly switches between negative and positive one in the topological regime, which indicates a finite-size fermion parity switch~\cite{hegde2016majorana}. Therefore, the Pfaffian of the Hamiltonian is only useful to define the topology of the infinite (or closed system, i.e., periodic). For the finite system, we refer to the LE as shown in the blue line in Fig.~\ref{fig:BDIPf} of the Appendix, where TQPT occurs when LE crosses zero.
	
	Next, we turn on a weak disorder $ \sigma_\mu/\bar{\mu} =0.4 $, as shown in the second column of Fig.~\ref{fig:BDI}. We find that the TQPT is slightly shifted due to disorder. In the trivial regime,  the gap of bulk states decreases to around $ 0.2\Delta $, which is much lower than that in the pristine limit. Similarly, in the topological regime, the gap of bulk states is also much smaller due to a disorder effect. Although some degrees of disorder show up in this situation, the topological properties of the system are still mostly preserved--- there is no trivial zero-energy bound state, and the topological zero-energy bound state is still protected by a finite topological gap above TQPT. This directly demonstrates the stability of the topological Majorana mode to weak, but finite, disorder.  Weak disorder does not destroy topological superconductivity or the Majorana zero modes, and ``weak'' here is not that small since it is 40\% of the nominal chemical potential.
	
	Then we keep increasing disorder to an intermediate level $ \sigma_\mu/\bar{\mu} =1 $ as shown in the third column of Fig.~\ref{fig:BDI}. We find that the gap of the bulk state completely collapses in the trivial regime, and does not reopen in the topological regime as shown in Fig.~\ref{fig:BDI}(g). This is similar to most existing transport experiments in the nanowire, where the signatures of the gap closure and reopening features are both inconclusive. In the second and third row [Figs.~\ref{fig:BDI}(h) and (i)], we also show that the end-to-end correlation disappears due to disorder, which is also similar to the conductance behavior of ``ugly'' trivial zero peaks in the disordered nanowire~\cite{pan2021threeterminal}. In Fig.~\ref{fig:BDI}(h), the zero-energy bound states only exist in the trivial regime, and nothing in the topological regime, which means that all the zero-energy bound states are trivially induced by disorder, as for the ``ugly'' peaks in Ref.~\onlinecite{pan2020physical}. We also note that the effective TQPT now is no longer a point and extends over the whole hatched region. This is because the LE is not stable around zero, as shown in Fig.~\ref{fig:BDIPf}(c) of the Appendix. The LE first crosses zero at around $ t=2.5\Delta $ and then goes up. It crosses zero again near $ t=4.5\Delta $ before completely entering the negative region. Therefore, the first time it crosses zero heralds the beginning of the transition from the trivial to the topological regime, and the last time it crosses zero defines the completion of the transition. The distinction between topological and trivial is no longer sharp in the sense of a phase transition in the presence of strong disorder.
	
	Finally, we show the results for the Kitaev chain in the presence of a large disorder $ \sigma_\mu/\bar{\mu}=5 $ in the fourth column of Fig.~\ref{fig:BDI}. In this very large disorder, the system enters the regime of Anderson localization, where everything becomes featureless. There is nearly no stable zero-energy bound state lying on the axis of $t $. In this case, the Kitaev chain is always in the trivial regime because the LE is always positive as shown in Fig.~\ref{fig:BDIPf}(d). The topological regime has been replaced by a disorder-localized regime.

%\FloatBarrier
	\subsection{Short Kitaev chains in class D} \label{sec:D}
	\begin{figure*}[ht]
		\centering
		\includegraphics[width=\textwidth]{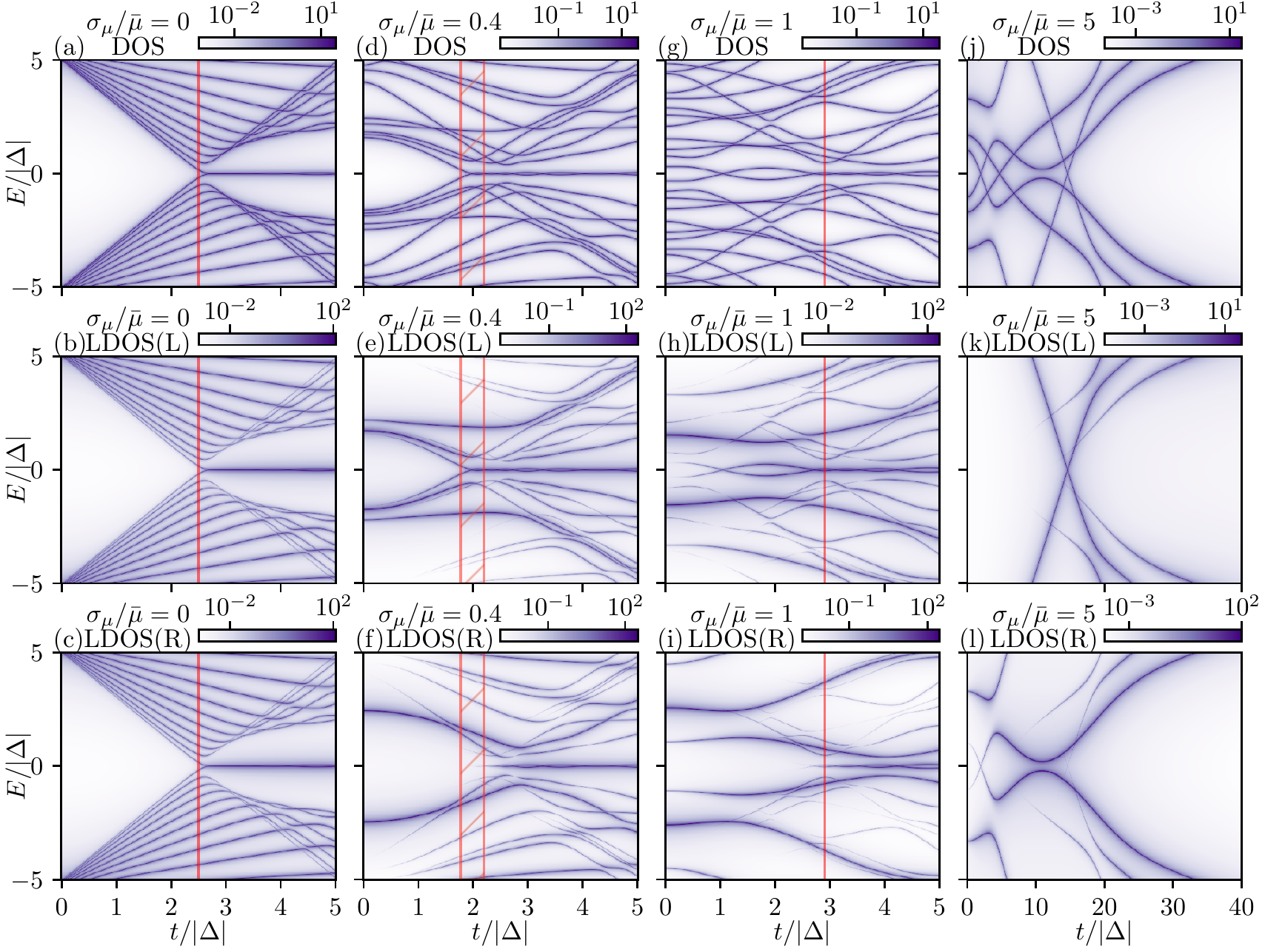}
		\caption{The representative results of a short Kitaev chain with $ L=20$ in class D ($ \theta=\frac{\pi}{2} $). 
		(a)-(c) show the pristine wire; 
		(d)-(f) show a weak disorder with $ \sigma_\mu/\bar{\mu}=0.4 $; 
		(g)-(i) show an intermediate disorder with $ \sigma_\mu/\bar{\mu}=1 $;  
		(j)-(l) show a strong disorder with $ \sigma_\mu/\bar{\mu}=5 $. The first row represents the DOS; the second row represents the LDOS at the left end; the third row represents the LDOS at the right end. Refer to Fig.~\ref{fig:BDI} for the definitions of red lines, hatched regions, and other parameters. The corresponding LE and Pfaffian are shown in Fig.~\ref{fig:DPf}.}
		\label{fig:D}
	\end{figure*}
	
	For a direct comparison, we break the time-reversal symmetry by setting the phase of $ \Delta $ to $ \frac{\pi}{2} $, and calculate the short Kitaev chain in class D again as shown in Fig.~\ref{fig:D}. The first column shows the pristine case, where we find an identical result to the Kitaev chain in class BDI. This is expected since the phase of $ \Delta $ can be absorbed into Majorana operators. In the remaining three columns, we calculate the results in the presence of a weak disorder $ \sigma_\mu/\bar{\mu}=0.4 $, an intermediate disorder $ \sigma_\mu/\bar{\mu}=1 $, and a strong disorder $ \sigma_\mu/\bar{\mu}=5 $. Their results are all similar to the Kitaev chain in class BDI: Weak disorder still induces good Majorana zero modes, and good Majorana zero modes transmute into ugly zero modes as disorder increases. Finally, strong disorder does not induce any zero-energy bound state indicating complete localization.

	\subsection{Long Kitaev chains in class BDI} \label{sec:BDI50}
	\begin{figure*}[ht]
		\centering
		\includegraphics[width=\textwidth]{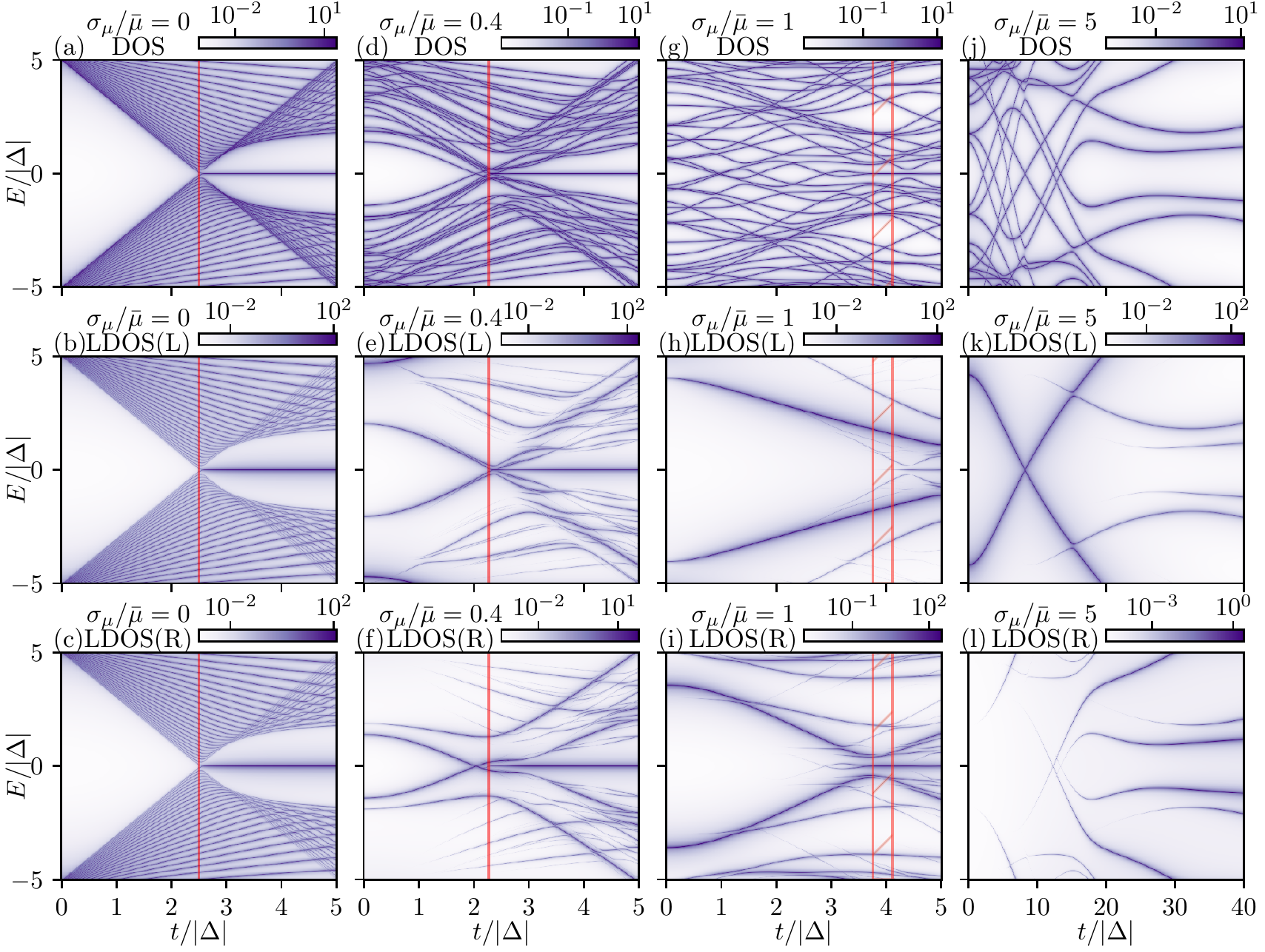}
		\caption{The representative results of a long Kitaev chain with $ L=50$ in class BDI ($ \theta=0 $). 
		(a)-(c) show the pristine wire; 
		(d)-(f) show a weak disorder with $ \sigma_\mu/\bar{\mu}=0.4 $; 
		(g)-(i) show an intermediate disorder with $ \sigma_\mu/\bar{\mu}=1 $;  
		(j)-(l) show a strong disorder with $ \sigma_\mu/\bar{\mu}=5 $.
		The first row represents the DOS; the second row represents the LDOS at the left end; the third row represents the LDOS at the right end. Refer to Fig.~\ref{fig:BDI} for the definitions of red lines, hatched regions, and other parameters. The corresponding LE and Pfaffian are shown in Fig.~\ref{fig:BDI50Pf}.}
		\label{fig:BDI50}
	\end{figure*}

	Next, we present a long Kitaev chain in class BDI in Fig.~\ref{fig:BDI50}. We extend the number of sites from $ L=20 $ (shown in Fig.~\ref{fig:BDI}) to $ L=50 $ (shown in Fig.~\ref{fig:BDI50}), and the effective wire length is therefore almost tripled. In the first column, we find that oscillations of zero-energy bound states are much suppressed because of the long wire limit. In addition, the TQPT is closer to the putative $ \frac{\mu}{2} $ which is for the infinite Kitaev chain.
	
	In the second column of Fig.~\ref{fig:BDI50}, we increase disorder to $ \sigma_\mu/\bar{\mu}=0.4 $, and find that the TQPT is almost unaffected by disorder. The topological properties are immune to weak disorder and preserved here. This corresponds to the good Majorana zero modes~\cite{pan2020physical} in the presence of weak disorder.
		
	In the third column of Fig.~\ref{fig:BDI50}, we increase disorder to an intermediate level $ \sigma_\mu/\bar{\mu}=1 $. The gap of bulk states is completely suppressed in the trivial regime, and the gap reopening feature is also absent. In this case, the TQPT, which is defined by calculating the LE, is already deviated from the putative TQPT due to disorder. Although the zero-energy bound states emerge above the TQPT, they are not protected by a finite topological gap, and thus are not robust against disorder.
	
	In the fourth column of Fig.~\ref{fig:BDI50}, we show that the system enters the Anderson localization regime in strong disorder, where the Kitaev chain is completely in the trivial regime and everything is featureless. There is no zero-energy bound state in this case.
%\FloatBarrier
	\subsection{Short superconductor-semiconductor nanowires} \label{sec:nw}
	\begin{figure*}[ht]
		\centering
		\includegraphics[width=\textwidth]{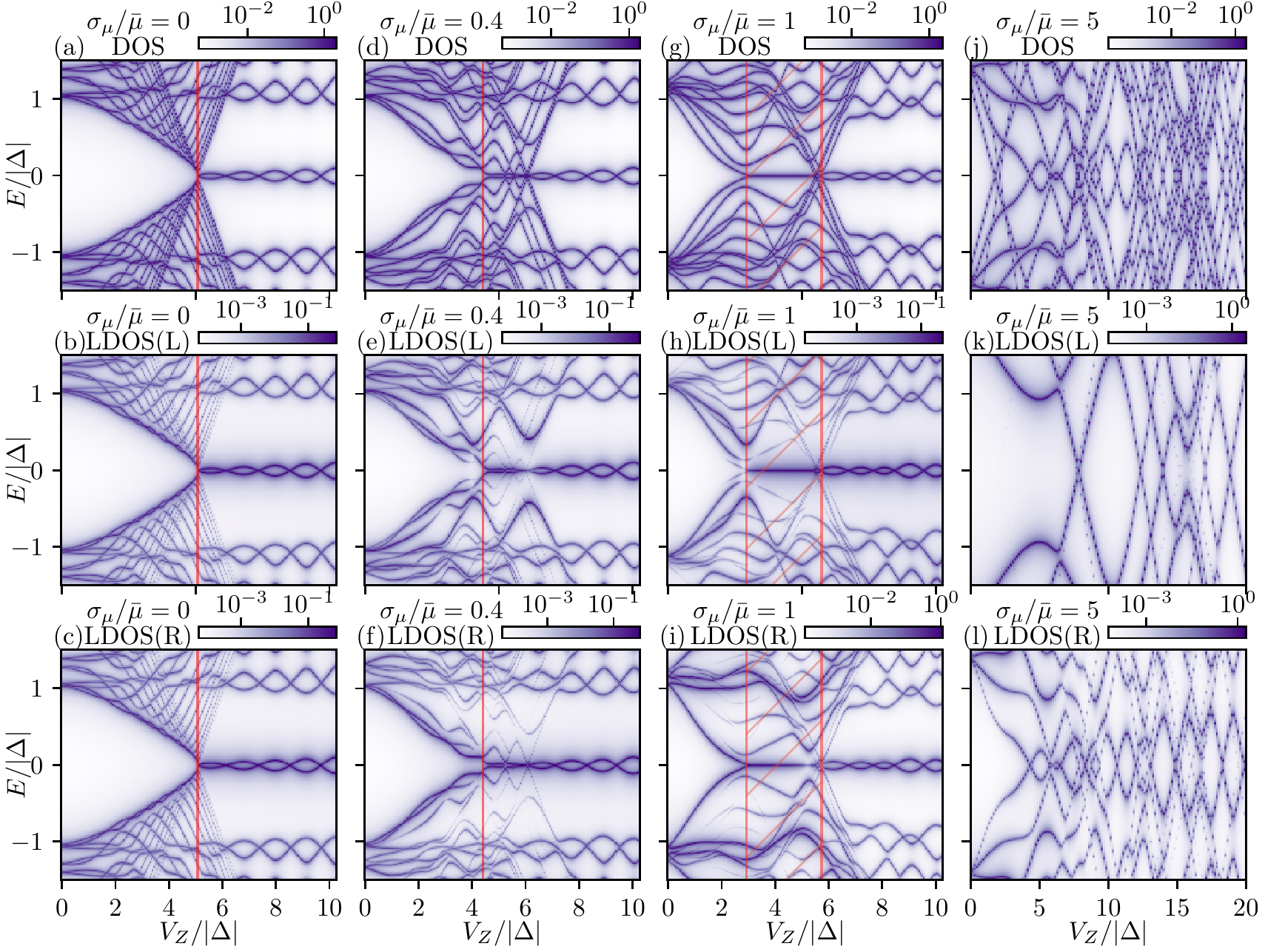}
		\caption{The representative results of a short nanowire with $ L=3~\mu$ m. 
		(a)-(c) show the pristine wire; 
		(d)-(f) show a weak disorder with $ \sigma_\mu/\bar{\mu}=0.4 $; 
		(g)-(i) show an intermediate disorder with $ \sigma_\mu/\bar{\mu}=1 $;  
		(j)-(l) show a strong disorder with $ \sigma_\mu/\bar{\mu}=5 $. The first row represents the DOS; the second row represents the LDOS at the left end; the third row represents the LDOS at the right end. The red line indicates the TQPT, which is defined by the TV. The hatched region represents a transition from the trivial to the topological regime, between which the TV is unstable around zero. The corresponding TV is shown in Fig.~\ref{fig:nwPf}. The other parameters are $ \bar{\mu}=1 $ meV, $ \Delta=0.2 $ meV, and $ \alpha=0.5  $ eV \AA.}
		\label{fig:nw}
	\end{figure*}
	To connect the the result of Kitaev chain to the realistic model, the nanowire, we now calculate the DOS and LDOS of the nanowire in the presence of disorder. First, we present the results of the short nanowire ($ 3~\mu $ m)  in Fig.~\ref{fig:nw} to compare with the short Kitaev chain as shown in Fig.~\ref{fig:BDI}. 
	
	In the first column of Fig.~\ref{fig:nw}, we first plot the pristine case as a reminder. The TQPT is defined as $ V_Z=\sqrt{\Delta^2+\mu^2}\sim 5.1\Delta $. In the pristine case, the gap closure and reopening features are visible with the gap closing exactly at the TQPT. The zero-energy Majorana bound states appear above the TQPT with an increasing Majorana oscillation~\cite{dassarma2012splitting}.
	
	In the second column of Fig.~\ref{fig:nw}, we introduce a weak disorder $ \sigma_\mu/\bar{\mu}=0.4 $. We find that although TQPT is slightly changed by disorder, the topological properties of the nanowire are still preserved: The gap closure and reopening features are visible and all the zero-energy bound states here are topological Majorana zero modes.  This is an example of good Majorana zero modes~\cite{pan2020physical} in the presence of weak disorder. We note again that disorder immunity applies to rather large disorder, almost 50\% of the chemical potential.

	However, when disorder increases to an intermediate level, as shown in the third column of Fig.~\ref{fig:nw}, we find that TQPT is strongly affected by disorder. Although the topological zero-energy bound states are still appearing beyond the TQPT, the disorder-induced trivial zero-energy bound states now emerge below the TQPT. These trivial states correspond to the ugly zero modes~\cite{pan2020physical} in the tunneling experiment that measures the conductance.
	
	Lastly, we tune disorder to a strong magnitude, $ \sigma_\mu/\bar{\mu}=5$. Similar to all the strong disorder cases from Figs.~\ref{fig:BDI} to~\ref{fig:BDI50}, there are just some random states crossing the zero-energy axis, with almost no occurrences of zero-energy bound states. Strong disorder manifests rather featureless behavior in DOS and LDOS as everything is now localized in the spectrum.

%\FloatBarrier
\subsection{Long superconductor-semiconductor nanowires} \label{sec:nw1000}
\begin{figure*}[ht]
	\centering
	\includegraphics[width=\textwidth]{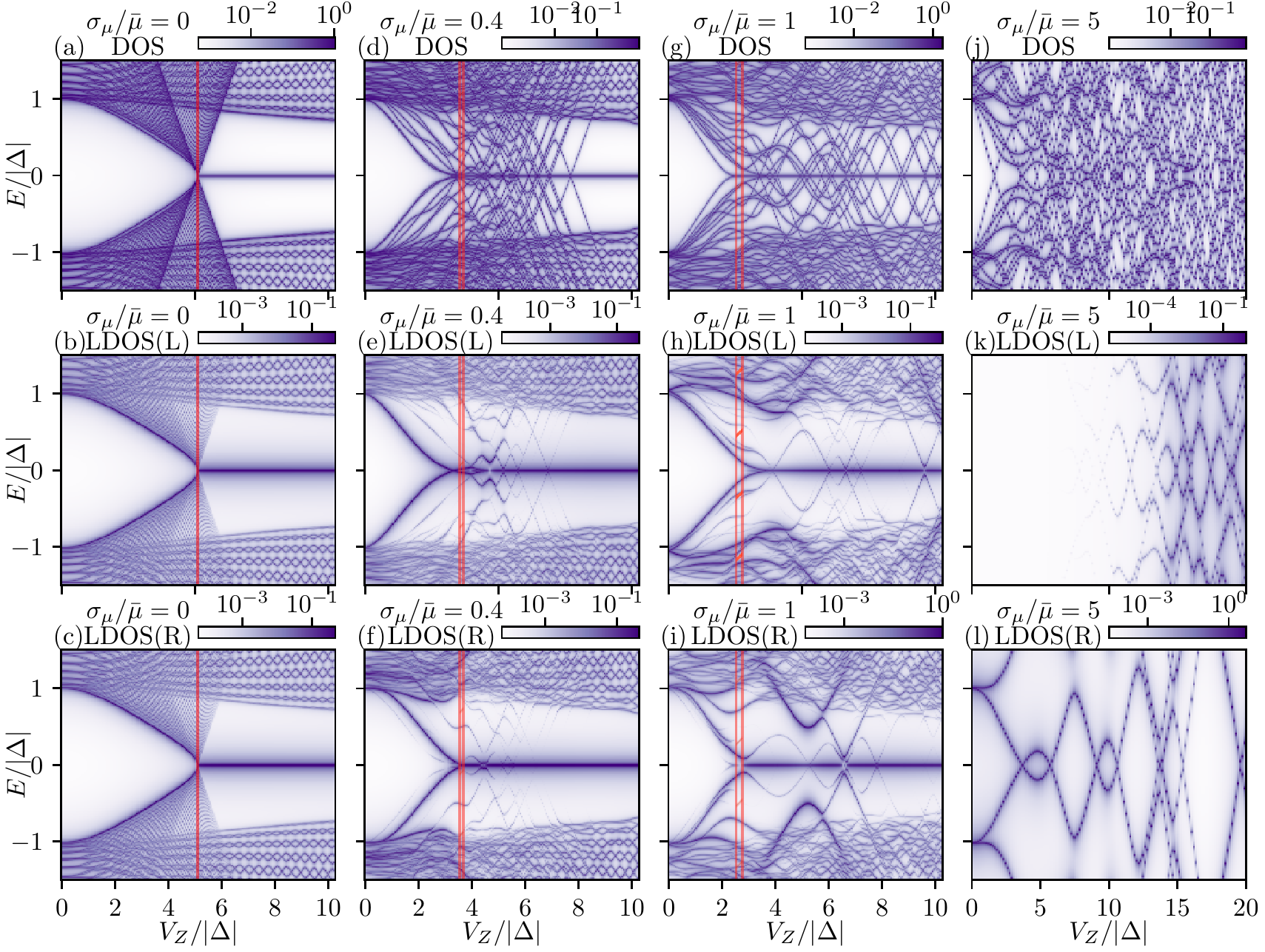}
	\caption{The representative results of a long nanowire with $ L=10~\mu$ m. 
	(a)-(c) show the pristine wire; 
	(d)-(f) show a weak disorder with $ \sigma_\mu/\bar{\mu}=0.4 $; 
	(g)-(i) show an intermediate disorder with $ \sigma_\mu/\bar{\mu}=1 $;  
	(j)-(l) show a strong disorder with $ \sigma_\mu/\bar{\mu}=5 $. The first row represents the DOS; the second row represents the LDOS at the left end; the third row represents the LDOS at the right end. The corresponding TV is shown in Fig.~\ref{fig:nw1000Pf}. Refer to Fig.~\ref{fig:nw} for the definitions of red lines, hatched regions, and the other parameters.}
	\label{fig:nw1000}
\end{figure*}

 	Finally, we present the last situation by replacing the short nanowire with a long one ($ L=10~\mu $ m) as shown in Fig.~\ref{fig:nw1000}, which should be compared with the Kitaev chain of 50 sites in Fig.~\ref{fig:BDI50}. 
 	
 	In the first column of Fig.~\ref{fig:nw1000}, we present the pristine long nanowire. This is almost the same as the short nanowire, except for the expected suppressed Majorana oscillation beyond TQPT.
 	
 	Then we apply a weak disorder to the long nanowire, as shown in the second column of Fig.~\ref{fig:nw1000}. Disorder does not play an important role here since the long wire is robust against weak disorder. The Majorana zero modes appear above the TQPT, and no zero-energy bound states are induced in the trivial regime. This is again an example of the disorder immunity of good Majorana zero modes in the presence of weak disorder.
 	
 	Next, we increase disorder to an intermediate level as shown in the third column of Fig.~\ref{fig:nw1000}. We find that the nanowire undergoes a continuous transition from the good to ugly zero modes. In the trivial regime, the absence of zero-energy bound states ensures that all the zero-energy states are Majorana zero modes, which is benign. However, in the topological regime, the zero-energy bound state beyond TQPT is not well protected by a finite topological gap. Although the gap closure feature is still very prominent, we do not see a gap reopening feature because too many bulk states cross zero in the topological regime~\cite{huang2018metamorphosis}.
 	
 	In the last column of Fig.~\ref{fig:nw1000}, we increase disorder to a very strong limit, and everything becomes completely random again and no characteristic features appear. The system is always in the trivial regime without any occurrence of zero-energy bound states in this Anderson localized strong disorder regime. 

\section{Discussion}\label{sec:discussion}

In this section, we discuss the disorder ensemble statistics of the zero-energy bound states in both models based on the five cases presented before: the short Kitaev chain in class BDI (BDI-20), the short Kitaev chain in class D (D-20), the long Kitaev chain BDI (BDI-50), the short nanowire (NW-3$ \mu $ m), and the long nanowire (NW-10$ \mu $ m). Each ensemble uses 100 random disorder configurations for averaging. 

Then we tune disorder from zero to a very large magnitude $ \sigma_\mu/\bar{\mu}=5 $, and extract the absolute energy interval of trivial zero-energy bound states. We measure this absolute energy interval on the axis of $ t $ for Kitaev chains and the axis of Zeeman field $  V_Z $ for nanowires. By taking the ensemble average, we can roughly estimate the likelihood of the occurrence of trivial zero-energy bound states in each case. 

However, since different ensembles may have different TQPTs, the absolute energy interval may not reflect the real likelihood of trivial zero-energy bound states. For example, given the same probability of the occurrence of trivial zero-energy bound states, if the TQPT in one ensemble is higher in energy than the other, then the average absolute energy interval of the trivial zero-energy states in that ensemble must be larger than the other. Such situations can happen in different ensembles with different disorder strengths because the large disorder can strongly affect the TQPT [e.g, Fig.~\ref{fig:D}(g)]. 

\begin{figure}[htbp]
	\centering
	\includegraphics[width=3.4in]{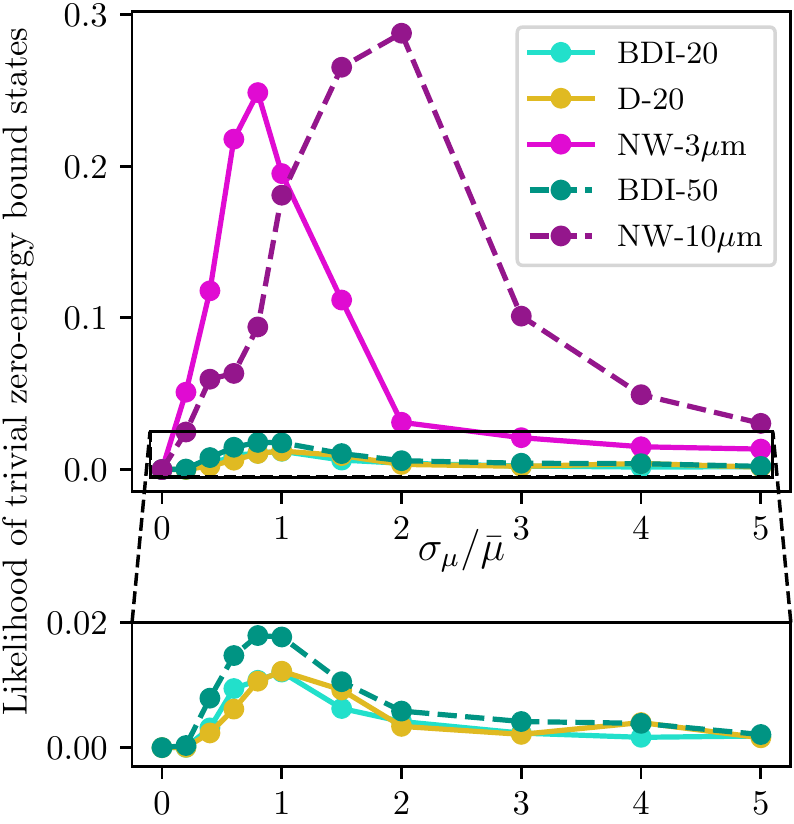}
	\caption{The likelihood of trivial zero-energy bound states as a function of disorder strength from weak ($ \sigma_\mu/\bar{\mu}=0.2 $) to strong ($ \sigma_\mu/\bar{\mu}=5 $) disorder. The likelihood is defined as the ensemble average of the ratio of the absolute energy interval of the trivial zero-energy bound states to the energy interval of the trivial regime, {where the ensemble size is 100}. The short wires are in solid lines ($ L=20 $ for the Kitaev chain and $ L=3~\mu $ m for the nanowire) while the long wires are in dashed lines ($ L=50 $ for the Kitaev chain and $ L=10~\mu $ m for the nanowire). In general, the likelihood shows a dome-shaped dependence on disorder strength. The nanowire is more susceptible to disorder than the Kitaev chain. The inset at the bottom shows the likelihood of the Kitaev chain in a higher resolution.}
	\label{fig:stat}
\end{figure}

Thus, we normalize the absolute energy interval of the trivial zero-energy bound states by the energy interval of the trivial regime, and then take the ensemble average to define the likelihood of the occurrence of trivial zero-energy bound states. This likelihood varies between zero and one, where zero corresponds to the pristine case without the occurrence of any trivial zero-energy bound states.  The results for the five cases are presented in Fig.~\ref{fig:stat}, with the inset at the bottom showing the three Kitaev chain ensembles in higher resolution.

First, we study the general behavior of this likelihood as a function of disorder strength. We find that the statistical likelihood of the occurrence of disorder-induced trivial zero-energy bound states in both the Kitaev chain and the nanowire ensemble has a universal dome-shaped dependence on the strength of disorder with the peak for the occurrence of disorder-induced trivial zero modes always occuring at intermediate disorder of the order of the chemical potential itself. 

In the weak disorder regime ($ \sigma_\mu/\bar{\mu}\lesssim 1 $), the likelihood of inducing trivial zero-energy bound states increases as disorder increases. This is because both systems maintain some degrees of robustness against disorder, but this protection is not infinite. The robustness against disorder gradually becomes weaker as disorder increases.

As the likelihood reaches its maximum at an intermediate disorder, the trivial zero-energy bound states are most likely to exist in the system. However, as disorder increases further to the strong disorder regime, the likelihood of the trivial zero-energy bound states declines again as the system enters the Anderson localization regime, which is essentially featureless. Thus, generically, disorder induces trivial zero modes mostly in the intermediate disorder regime, and the strong disorder regime is of course by definition entirely trivial since the system is now localized.  The weak disorder regime (disorder strength $ < $ half the chemical potential) is reasonably protected against disorder effects.

Second, we compare different ensembles to make a distinction in the role of disorder. The first comparison is made between the Kitaev chain and the nanowire, which is also our key finding. We find that the nanowire is much more susceptible to disorder than the Kitaev chain. The trivial zero-energy bound states are much more likely to emerge in the nanowire. The fact that disorder effects are an order of magnitude weaker (in the sense described above) in the Kitaev chain than in the nanowire may have contributed to the lack of appreciation by the community on the key significance of disorder in the interpretation of the nanowire Majorana tunnel conductance data until very recently~\cite{pan2020physical,pan2020generic}, in spite of some earlier works specifically pointing out the possibility of class D antilocalization peaks possibly masquerading as Majorana conductance peaks in the strong disorder regime~\cite{liu2012zerobias,sau2013density,mi2014xshaped}. As can be seen in Fig.~\ref{fig:stat}, although the universal behaviors of disorder-induced trivial peak statistics are similar for all cases in the weak and strong disorder limits, the nonuniversal behaviors in the intermediate disorder regime are very different in Kitaev chains and nanowires.

The second comparison is made within the Kitaev chain, which is presented in the inset of Fig.~\ref{fig:stat}. We first find that there is no apparent distinction between the statistics of the Kitaev chain in class BDI and class D, because they can be transformed into each other by redefining Majorana operators~\cite{kitaev2001unpaired}. Thus, disorder does not play any different role in class D or class BDI. 

We also compare the short Kitaev chain and long Kitaev chain, and find that the longer chain does not provide significant extra protection against disorder--- the likelihood curve (the cyan dashed line) in the long Kitaev chain is qualitatively the same as those in the short Kitaev chains (turquoise and yellow solid lines). This is because the short Kitaev chain is already rather robust against disorder.

However, this situation is different when it comes to the long nanowire. As we mentioned before, the nanowire is more susceptible to disorder. In Fig.~\ref{fig:stat}, we find that the likelihood of trivial zero-energy bound states in the short nanowire (shown in the magenta solid line) is maximal at around $ \sigma_\mu/\bar{\mu}=1 $  while it reaches its peak at around $ \sigma_\mu/\bar{\mu}=2 $ in the long nanowire (shown in the purple dashed line). This implies that the short nanowire tends to enter the Anderson localization regime sooner than the long nanowire as we increase the disorder strength from zero. In other words, the ``weak" disorder regime in the long wire is larger than that in the short wire. Thus, we expect to see more good Majorana zero modes in the long nanowire than in the short nanowire. In addition, before they both enter the Anderson localization regime ($ \sigma_\mu/\bar{\mu}\lesssim 1 $), at a given disorder strength, the long nanowire always has a smaller likelihood of trivial zero-energy bound states than the short nanowire. Thus, future experiments should not only focus on reducing the actual disorder in the experimental samples, but should also try to use longer wires so that the effective topological immunity to disorder is substantially enhanced compared with the shorter wire situation.  Our work shows that even for the same dimensionless disorder strength relative to the chemical potential, longer wires will manifest many fewer trivial zero modes, making the Majorana observation more likely.
%Therefore, this manifests that, unlike the Kitaev chain, the nanowire will greatly benefit from the long wire limit to fight against disorder. 

\section{Conclusion}\label{sec:conclusion}

In this paper, we study the on-site disorder effect on a Kitaev chain and semiconductor nanowire, and calculate the LDOS in the presence of disorder in the aforementioned five cases: a short Kitaev chain in class BDI and class D, a long Kitaev chain in class BDI, a short nanowire, and a long nanowire. We compare the occurrence of trivial zero-energy bound states arising from disorder in these five distinct cases.

To define the TQPT in the finite system, we calculate (see the Appendix) the LE for the Kitaev chain and TV for the nanowire. We find that the real TQPT of finite systems can only be captured by the LE when it crosses zero since the Pfaffian of the Kitaev chain Hamiltonian can only represent the fermion parity switch. In the nanowire, we use zeros of TV to define the TQPT. We also find that disorder renormalizes the TQPT considerably, making TQPT deviated from the theoretical value in the long wire and pristine limit. 

When disorder is weak (typically $ \sigma_\mu/\bar{\mu} \lesssim 1 $), the bulk gap below and above the TQPT gradually decreases. However, the topological properties are still preserved before the gap completely collapses. In this weak disorder regime, the topological zero-energy bound states are still robust above TQPT while the trivial zero-energy bound states begin to emerge in the trivial regime as disorder increases. A clear distinction is possible between topological and trivial in this weak disorder regime.

The occurrence of trivial zero-energy bound states reaches its maximum when disorder goes up to $ \sigma_\mu/\bar{\mu} \sim 1 $ for the short wire situation. Topological properties are not preserved, and the topological zero-energy bound state is not protected by a finite topological gap. In this intermediate disorder situation, the likelihood of the zero modes being of a trivial origin is high.

In the strong disorder regime, we find that the likelihood of the emergence of trivial zero-energy bound states decreases. There are only random states contributing to the LDOS as a result of Anderson localization. Thus, it is not meaningful to discuss the Kitaev chain or nanowire here because the system is completely dominated by disorder, which should be better described by a random matrix approach~\cite{beenakker1997randommatrix,guhr1998randommatrix,brouwer1999distribution,beenakker2015randommatrix,mi2014xshaped,pan2020generic}.

Our key finding is that, although disorder affects both the Kitaev chain and nanowire, its effect is highly nonuniversal. In general, the disorder-induced trivial zero-energy bound states in the nanowire are much more prominent than in the Kitaev chain. Only in the regimes of very weak and very strong disorder do universal features emerge, but experimentally the relevant regime is likely to be a generic intermediate disorder regime where the nanowire is much more susceptible to disorder than the Kitaev chain.

We also consider the long wire situation comparing between the long Kitaev chain and the long nanowire. We find that the nanowire benefits from the long wire limit more than does the Kitaev chain. In particular, for long nanowires the topological immunity persists to a much higher dimensionless disorder strength compared with the corresponding Kitaev chain situation.

In summary, we have compared the effect of disorder on the Kitaev chain and the semiconductor nanowire with respect to their topological immunity and the occurrence of disorder-induced trivial zero modes.  Our main finding is that although disorder effects in the two systems are in general similar, the nanowire is much more quantitatively susceptible to disorder effects than the Kitaev chain with the probability of a disorder-induced zero mode arising in the nanowire being much (by an order of magnitude at least) higher in the nanowire than in the Kitaev chain.  We find that increasing the nanowire length considerably enhances its topological immunity to disorder compared with the corresponding situation in a Kitaev chain. We also find that both systems have topological immunity for weak disorder, and the strong disorder regime is universal and featureless in both systems, as it is completely dominated by localization.

This work is supported by Laboratory for Physical Sciences and Microsoft. We also acknowledge the support of the University of Maryland supercomputing resources~\cite{hpcc}.

\bibliographystyle{apsrev4-2}
\bibliography{Paper_KitaevChain}

\appendix
\setcounter{secnumdepth}{3}
\renewcommand\figurename{Figure}
\onecolumngrid

\section{Pfaffian and Lyapunov exponents}\label{sec:appendix}
\begin{figure*}[htbp]
	\centering
	\includegraphics[width=6.8in]{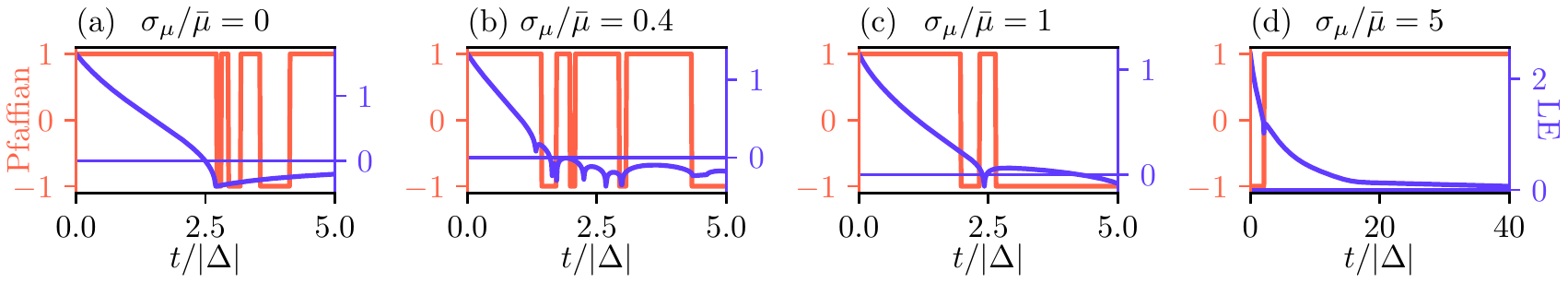}
	\caption{The corresponding Pfaffian and LE of the short Kitaev chain in class BDI with $ L= 20 $ in Fig.~\ref{fig:BDI}.}
	\label{fig:BDIPf}
\end{figure*}

\begin{figure*}[htbp]
	\centering
	\includegraphics[width=6.8in]{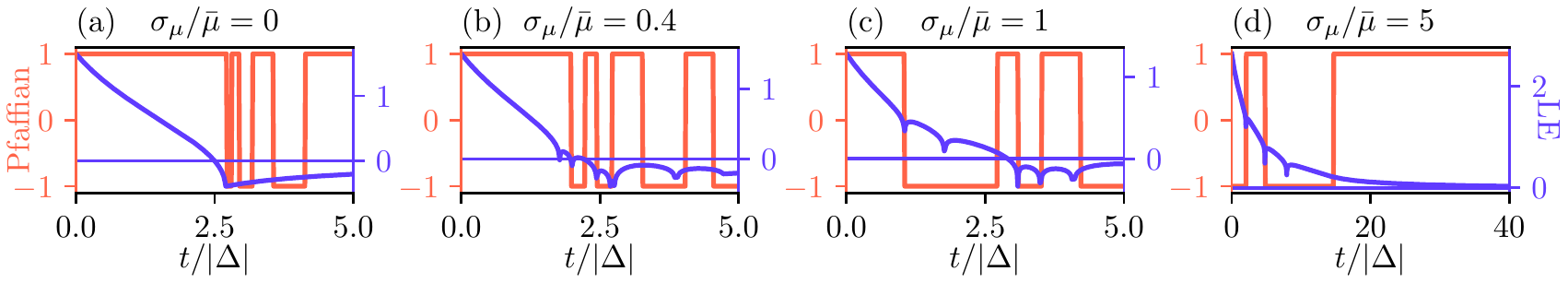}
	\caption{The corresponding Pfaffian and LE of the short Kitaev chain in class D with $ L= 20 $ in Fig.~\ref{fig:D}.}
	\label{fig:DPf}
\end{figure*}

\begin{figure*}[htbp]
	\centering
	\includegraphics[width=6.8in]{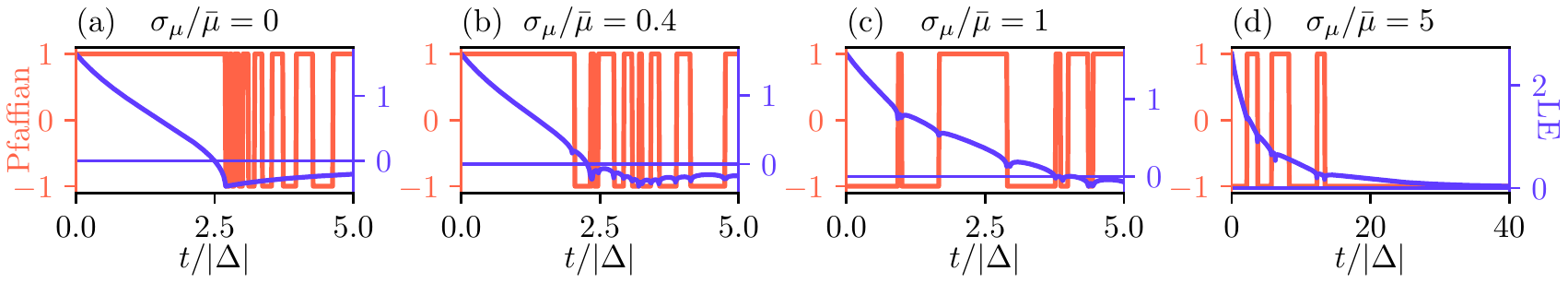}
	\caption{The corresponding Pfaffian and LE of the long Kitaev chain in class BDI $ L= 50 $ in Fig.~\ref{fig:BDI50}.}
	\label{fig:BDI50Pf}
\end{figure*}

\begin{figure*}[htbp]
	\centering
	\includegraphics[width=6.8in]{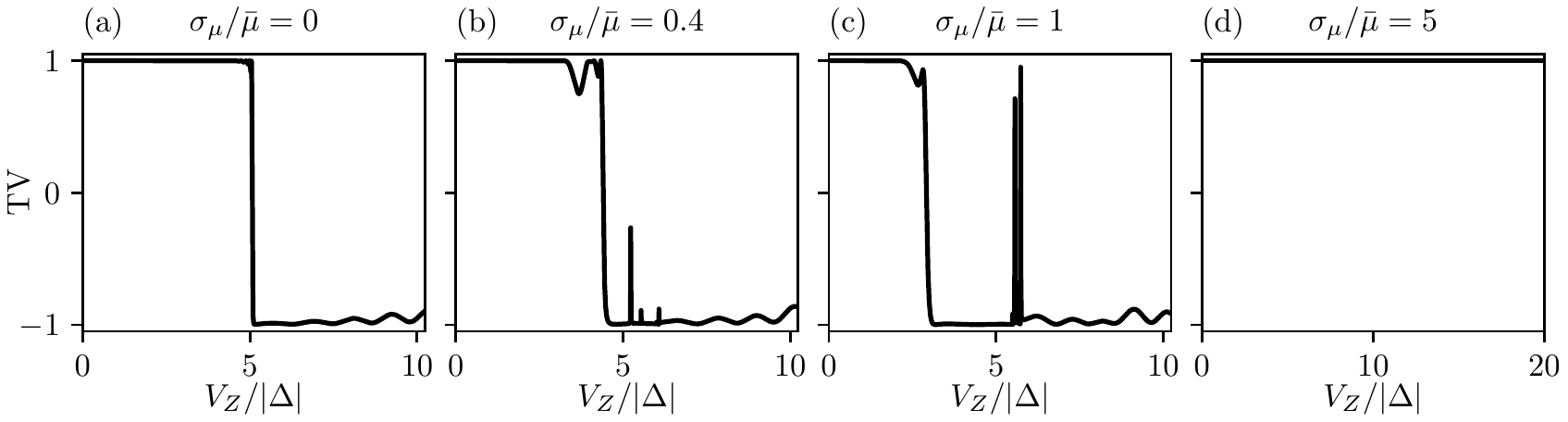}
	\caption{The corresponding TV of the short nanowire with $ L=3~\mu$ m in Fig.~\ref{fig:nw}.}
	\label{fig:nwPf}
\end{figure*}

\begin{figure*}[htbp]
	\centering
	\includegraphics[width=6.8in]{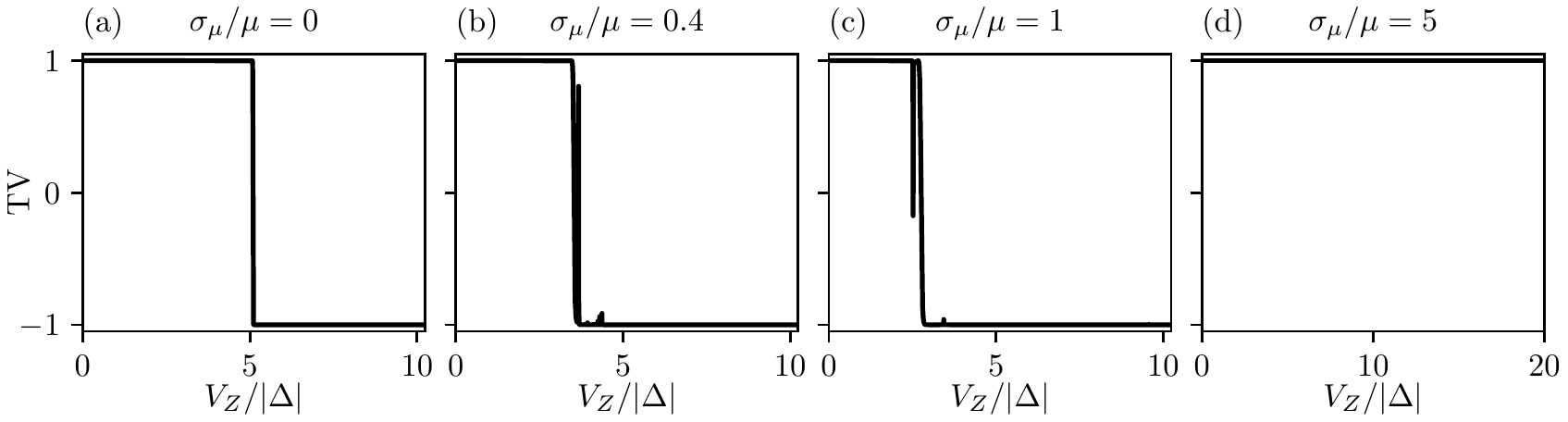}
	\caption{The corresponding TV of the long nanowire with $ L=10~\mu$ m in Fig.~\ref{fig:nw1000}.}
	\label{fig:nw1000Pf}
\end{figure*}

We present the calculated Pfaffian and LE corresponding to the results in the main text in Figs.~\ref{fig:BDIPf}-\ref{fig:BDI50Pf}, and the TV in Figs.~\ref{fig:nwPf} and~\ref{fig:nw1000Pf}.

\end{document}